# Atomistic study on cooling rate induced nanoindentation properties of Additively Manufactured Inconel-718

Toushiqul Islam, Md Samin Ashiq Aziz, Sadib Fardin, Abrar Faiyad, Mohammad Motalab*

*Department of Mechanical Engineering, Bangladesh University of Engineering and Technology, Dhaka - 1000, Bangladesh*

toushiqul.islam03@gmail.com, ashiqub@gmail.com, sadibfardin@gmail.com, afaiyad@ucmerced.edu, abdulmotalab@me.buet.ac.bd (*corresponding author)

## Abstract

Inconel-718's compatibility with additive manufacturing (AM) has made it a center of attention for researchers. This paper focuses on how cooling rates affect the hardness of AM Inconel-718. To study the AM process, monocrystalline and polycrystalline Inconel-718 layers were added to a pristine substrate and equilibrated at 2000 K before being cooled to 300 K using cooling rates ranging from 5 K/ps to 100 K/ps, as well as an exponential cooling rate. The layers were then subjected to atomistic nanoindentation simulation to analyze the nanomechanical response, including hardness, dislocation density, microstructure, and surface imprints, at different cooling rates. Load-displacement (P-h) curves were plotted for each cooling rate. The findings of this study provide crucial insights into the effect of cooling rate on the nanoindentation-based response of additively manufactured Inconel-718. These insights can aid in the design of high-performance components for various applications.

## Introduction

Extensive research and development have focused on Inconel-718 (IN-718), which is an austenitic superalloy based on Ni-Cr-Fe. Due to its high-temperature strength, Inconel-718 has been widely utilized as a core component in applications that require high temperatures, low thermal diffusivity, high corrosion resistance and fatigue, and wonderful weldability [1,2]. For the past several decades, it has been used in various critical applications, including 3-dimensional metal-based printing, die casting, components of gas turbines, turbocharger rotors, reactors in nuclear power plant, rockets, robotic vehicles, the aerospace industry, and other civil engineering applications [1,3]. IN-718 is greatly valued in sectors that prioritize weight savings due to its magnificent mechanical properties, leading to extensive research aimed at understanding its behavior and the underlying reasons behind its' failure.

The use of additive manufacturing (AM) is now common in the industry for creating metal parts as it offers greater flexibility in part design. There are several AM processes in order to produce metal components, including vat photopolymerization (SLA) [4], fused filament fabrication (FFF) [5], laminated object manufacturing (LOM) [6], Powder Bed Fusion (PBF) [7], selective laser melting (SLM) [8], direct metal deposition (DMD) [9], laser engineered net shaping (LENS) [10], direct metal laser melting (DMLM) [11] along with many other methods. Among these, the most widely used techniques are powder bed fusion-based AM processes, such as SLA, SLM, LENS, or DMLM. Several processing parameters play a prominent role in determining the mechanisms and properties of the final product [1,12]. Wang et al. conducted a thorough research to find out the effects of laser scan rate as well as power on fabricated sample residual stresses [13]. Their findings indicate that elevating the laser power led to greater residual stress along the build direction, whereas adjusting the laser speed influenced stresses in the transverse directions of the additive samples.

Additive manufacturing (AM) has revitalized the interest in IN-718, that was previously restricted in its usefulness due to its hardness, low thermal conductivity, work hardening characteristics, and challenges with being machined. AM, on the other hand, enables efficient and precise manufacturing of complex-shaped unibody structures [1,4,12], which has made it possible to manufacture lightweight intricate shapes of IN-718, including very tiny system like nanowires, nanopillars, and thin plate. Traditional machining issues, such as high cutting forces and resulting metallurgical changes, have been overcome, making IN-718 an ideal material for AM. Consequently, it is possible to decrease the weight of produced components without compromising their durability [14], and interest in IN-718 has multiplied many times over. The nanoscale structure of IN-718 is a critical factor in determining the reliability of integral parts produced using AM. The existence of hierarchical structural characteristics such as voids, inclusions, precipitates, and grains in AM-produced IN-718 causes non-uniform distribution of strength throughout the component due to local variations in mechanical properties [2]. The way dislocations interact with these features is a critical factor in determining the type of failure that may occur. Post-processing techniques such as annealing and surface finishing are essential in considerably improving the mechanical properties of the components [15,16]. However, there is still incomplete comprehension of the function of dislocation density, voids, and cooling rates at the nanoscale in IN-718, as it is challenging to observe these phenomena experimentally. Molecular dynamics (MD) simulations are a computational tool that can be used to model dislocation density, quenching rates, and void-metal interactions [17,18]. MD simulations have made significant contributions to increase our knowledge of physics, including creep cavitation [19], monotonic and cyclic stress-strain behaviour [20], distribution of recast layer, and discharge properties [21]. These techniques have effectively been employed to gauge the density of IN-718 in both regular and metastable liquid phases, as well as to investigate the dislocation density of shock-compressed single-crystal Cu and tantalum [17,18] and the contribution of dislocation density in case of void interactions.

Extensive research has been conducted on indentation, which is a well-established method for determining the material properties of large-sized materials. However, when dealing with very small sample sizes, traditional indentation techniques are inadequate for predicting material properties. In the case of samples ranging from a few nanometers to micrometers in size, conducting traditional indentation experiments becomes impractical. Nevertheless, over the past few decades, there has been growing interest in nanoscale materials and devices, leading to the development of new indentation techniques tailored for small length scales [22]. Nanoindentation, a technique that has long been studied at the atomic level, has gained significant attention. Landman et al. utilized a molecular dynamics approach to investigate nanoindentation on a gold surface, employing nickel as the indenter [23]. Their study suggested that the atomistic approach is effective in measuring material properties and identifying the onset of plasticity. In their study, Szlufarska et al. explored the effects of nanoindentation on a SiC substrate and made an intriguing discovery. They observed that the material underwent a phase transformation while being subjected to loading, leading to its amorphization beneath the surface of the indenter [24]. Similar phase transformations were also observed in Si [25,26]. In their research, Ma and Yang conducted an analysis of nanoindentation on nanocrystalline Cu and compared it to single crystal Cu using the LJ potential [27] . They made a noteworthy finding that the occurrence and cessation of dislocation in nanocrystalline Cu significantly influence the plastic properties of materials. Li et al. carried out a comprehensive study on nanoindentation in Aluminum, combining experimental observations with molecular dynamics simulations, with indentation depths reaching up to 50 nm [28]. They visualized the pattern of dislocation bursts in their experimental results and employed molecular dynamics to explain the formation and propagation of dislocation loops under the applied load. Lee et al. investigated the dislocation patterns on the Al (111) surface, employing different types of interatomic potential. Their study shed light on the sites of nucleation, dislocation locks, and loop formation occurring just beneath the indenter tip, as well as the occurrence of prismatic dislocation loops away from the contact surface [29]. Wagner et al. studied the dislocation nucleation for single crystal Al using MD simulations at temperatures of 0−300K and proposed that temperature can pre-nucleate dislocations [30]. They also suggested that a smooth indenter nucleates dislocations below the contact surface, but a rough indenter can nucleate dislocations both at and beneath the surface. Begau et al. conducted a study on nanoindentation in Cu crystals, aiming to elucidate the generation of the first dislocation during the pop-in event and its subsequent multiplication through atomistic analysis [31]. Catoor et al. performed nanoindentation experiments on various surfaces of single-crystal magnesium (Mg) using a spherical indenter. Their research identified the slip system involved during failure and pop-in events [32]. Somekawa et al. explored nanoindentation in Mg through a combination of experimental and molecular dynamics (MD) approaches. Their investigation revealed that indentations on the basal plane exhibit higher pop-in load and displacement compared to those on the prismatic plane [33]. Ziegenhain et al. studied the initiation of plasticity and crystal anisotropy using nanoindentation techniques in both Al and Cu materials [34]. Their results suggested that crystal orientation primarily influences elastic deformation, whereas its impact on plastic deformation is relatively insignificant. Fu et al. examined the plasticity of Vn (001) crystals through nanoindentation, highlighting the role of partial dislocations in the initial stage of plasticity [35].

In nanoindentation experiments, various types of indenters are available, including conical, spherical, flat, cubic, cylindrical, Berkovich, among others. In molecular dynamics (MD) studies, the spherical indenter is commonly used. The selection of the indenter material and its modeling as an analytical rigid body or composed of hard materials like diamond or silicon carbide (SiC) is mainly driven by the ease of modeling and the opportunity to validate theoretical predictions. It is important to note that the shape of the indenter plays a crucial role in determining the indentation force and the hardness of the materials. This is because the interaction between the indenter and the substrate heavily influences the contact area. In MD studies, the indentation depth is often limited due to computational constraints. Typically, the indentation depth in MD simulations is varied up to a few nanometers, whereas experimental purposes can involve depths in the range of hundreds of nanometers, which are feasible in practice. Knap and Ortiz studied the indenter size effect for nanoindentation in Au(001) and analyzing the P −h diagram [36]. Verkhovtsev et al. conducted an MD simulation study on nanoindentation in titanium (Ti) crystal, investigating the effect of indenter shape. In this paper, a thorough atomistic study of additively manufactured single crystalline (pristine) and polycrystalline IN-718 microstructures was conducted under uniaxial tensile loading [37]. In order to gain a comprehensive understanding of the dynamic behavior of materials that contain defects such as dislocations and voids in nanostructures. These structures are known to be effective for analysis and provide an ideal test-bed for studying the behavior of materials at the nanoscale. The focus of this study is to investigate the influence of different cooling rates and dislocations that have undergone additive manufacturing process. Additionally, it explores the relationship between microstructures and the mechanical characteristics such as hardness of IN-718, and presents results on the failure behavior of IN-718, providing insights into the deformation mechanism at the nanoscale. By conducting thorough research, the study aims to develop our understanding of behavior of this material, and contribute to the improvement of efficient designs for additive manufacturing (AM) parts. Keeping the scopes in mind, this paper presents an atomistic study of a novel method of observing the microstructure evolution of IN-718 during AM applying unidirectional solidification principle under different cooling rates and substrate microstructure. To develop a detailed understanding of the residual stresses, dislocations, twins and slip planes formed during the AM process we have created a pristine and polycrystalline substrate and added molten layers of IN-718 above it. Then the molten layers are cooled to 300K at different cooling rates. A proper insight about the residual defects is obtained. Following this, a statistical analysis of the mechanical property of the constructed columns are conducted for a deeper understanding about the performance of the structures. This study can be helpful for understanding the microstructure evolution and the associated defect formation, and mechanical characteristics of additively manufactured IN-718.

## 2. Methodology:

## 2.1 Atomic Structure Modelling

Inconel 718 contains a complex composition of elements, including nickel, chromium, iron, molybdenum, niobium, titanium, aluminum, and other trace elements. Ni, Fe and Cr comprises of almost 91% of the alloy [38]. Modeling the potential energy surface of Inconel 718 with all its elements can be problematic due to the complexity of the interactions between the many elements. A study by wang et al [39] revelead that density function of simplified fcc structure of $Ni_{60}Cr_{21}Fe_{19}$ matches the characteristics of bulk inconel-718. So, the method used in wang et al's study is also followed in this paper.

Initially, a face-centered cubic (FCC) Ni crystal with a [0 0 1] orientation was created. Subsequently, a portion of the Ni atoms was substituted with Cr, accounting for approximately 21% of the total atoms. Additionally, Fe atoms were substituted for approximately 19% of the Ni atoms in the crystal structure. We used atomsk to create a 15 grained polycrystal using random grain method [39]. The layers that we wanted to deposit were constructed as cube with sides of 14.3 nm. The substrate where the layers were deposited was chosen to be a pristine structure with a square surface with sides of 14.3 nm and a height of 1.43 nm.

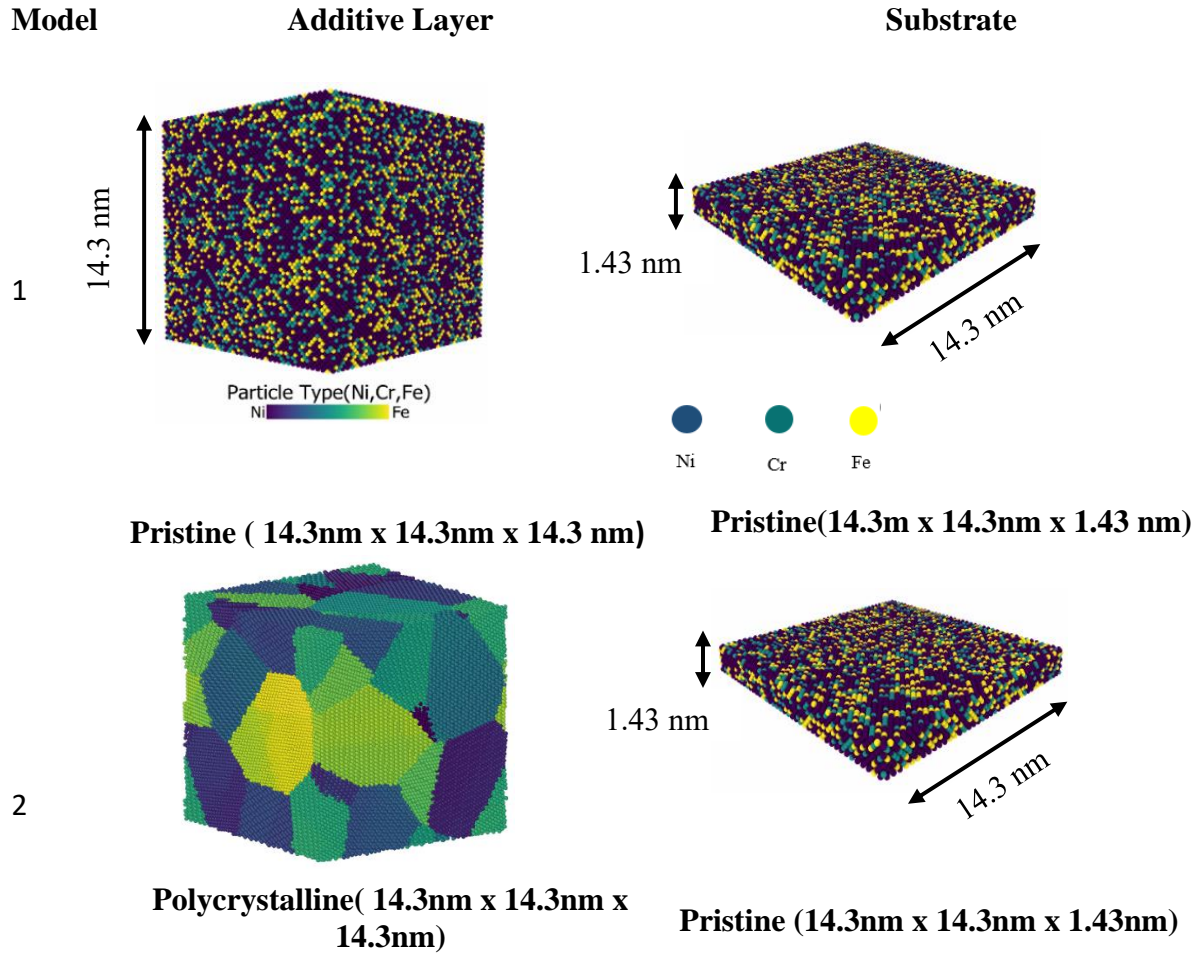

**Figure 2.1**: Models used in the present simulation

## 2.2 Additive Manufacturing

All simulations performed in this study are done using using large-scale atomic/molecular massively parallel simulator (LAMMPS) [40]. OVITO [41] is used for visualizing the microstructure evolution process along with dislocation loops during nanoindentation. In this work, we use the embedded atomic model (EAM) potential and parameters. The time step for AM process is 1 fs. We first melt the modeled additive layer and increased its temperature to 2000K in an NPT ensemble. The coordinates and energies of the molten layer are dumped for further simulation. For the AM simulation, both polycrystalline and pristine molten layer is added over the substrate and then equilibrated at 2000K. Following this the molten layer is allowed to cool from 2000 K to 300K at different cooling rtaes in an NPT ensemble. The different quenching profiles are presented in figure 2.3.

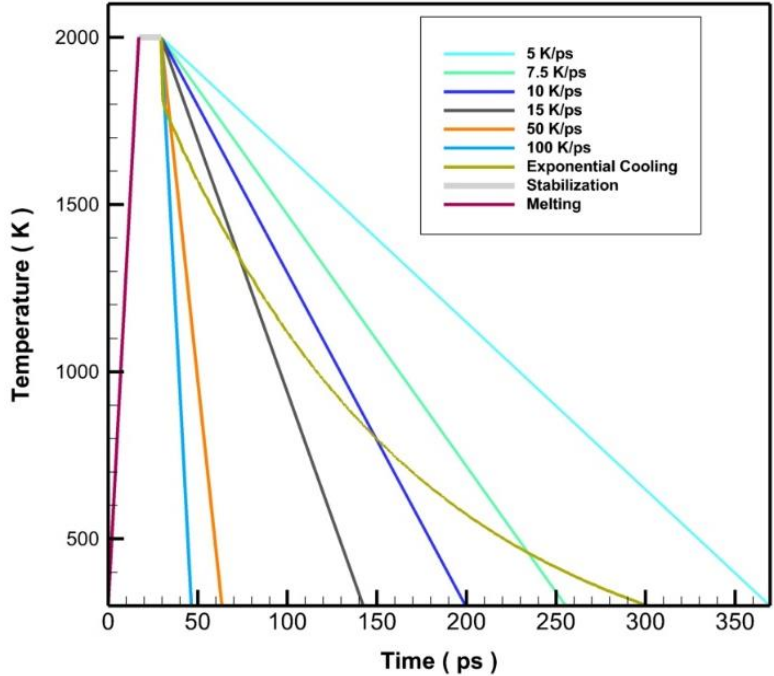

**Figure 2.3**: Quenching profile for various cooling rate

Melt spinning ($10^{-7}$–$10^{-6}$ K/ps)[42], liquid splat-quenching ($10^{-3}$ –$10^{-2}$ K/ps) [43], and pulsed laser quenching (1-10 K/ps) [44] are different quenching methods with various cooling rates. In this current study, cooling rates of 5K/ps, 7.5K/ps, and 10K/ps are used, which are akin to pulse laser quenching. Though higher cooling rates of 50K/ps, and 100K/ps are not achievable with current technologies; still they are considered in this study to predict the material behavior in such high cooling rates.

During quenching simulation the substrate was kept in an NVT ensemble with temperature set to 300 K. This makes the atoms near the substrate cool faster than the upper atoms and allows for a unidirectional solidification of the molten layer.

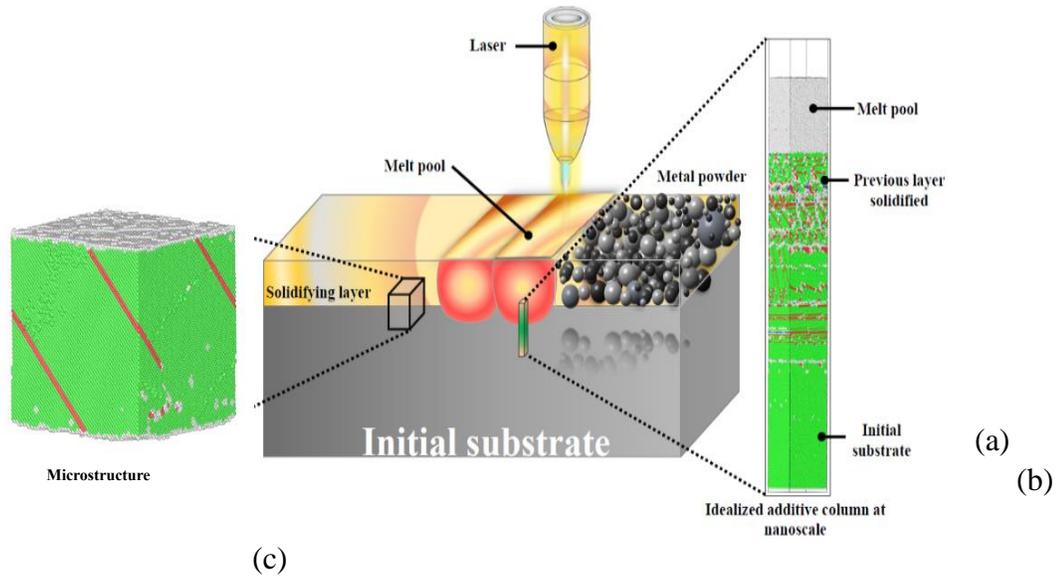

**Figure 2.4**: (a) Schematic AM process; (b) Idealized additive column at nanoscale;

(c) Solidified layer microstructure

This phenomenon is akin to the cooling which occurs during AM process. To observe the effect of microstructure of substrate we perform AM simulation on the the pristine and polycrystalline substrates using different cooling rates. These simulations allow us to get a more detailed understanding of the microstructure evolution that occurs due to the different cooling rate and the substrate microstructure used. A visual representation of our molecular modeling concept is shown in Figure 2.4. (b). Figure 2.4(c) depicts the Unidirectional solidification phenomenon and the layering method used for our work with the help of DXA.

## 2.2 Nanoindentation

After AM simulation the coordinates and energy are dumped for a quasi-static nanoindentation simulation. The simulation box was divided into two regions: a bottom region, where the atoms were fixed, and an upper region, where the indenter penetrated. The bottom fixed region served two purposes: it provided rigid support for the substrate, and it acted as a heat bath during the penetration of the indenter into the upper region. A rigid spherical indenter (virtual indenter in LAMMPS) was then placed over the substrate and the indenter was pushed into the material. (see Figure 2.5). The loading step was followed by an unloading step adopting displacement control of the indenter and the system was minimized using conjugate gradient method after every time step to maintain the quasi-static loading process at 0 K temperature.

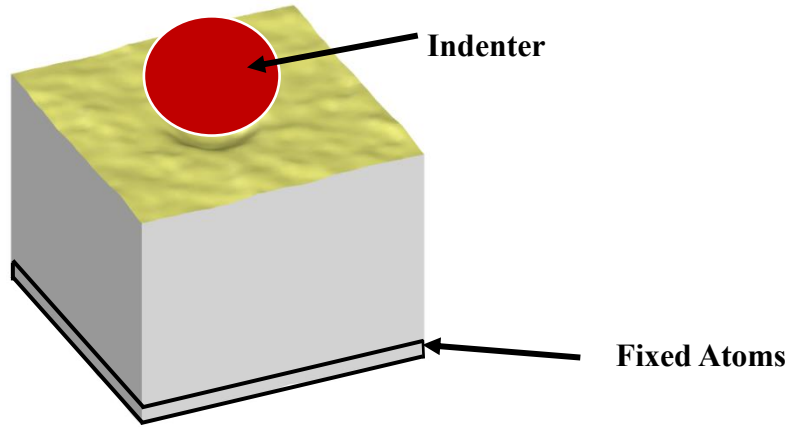

**Figure 2.5**: Schematic Diagram of Nanoindentation

| Indentation Speed | Indenter Radius | Indentation Depth | Indenter Force Constant |
|---|---|---|---|
| 50 m/s | 4 nm | 1.6 nm | $1\ eV/Å^3$ |

**Table 1**: Nanoindentation indenter parameters

## 2.3 Hardness Calculation

In order to calculate hardness, we used the Oliver-Pharr [45] and contact atom method [46].

### Oliver-Pharr Method:

A typical load-displacement curve is shown in Figure 2.6. The curve shows the relationship between the load applied to a material and the resulting displacement of the material. The load is increased up to a maximum value, $P_{max}$, and then unloaded. The unloading process creates a critical depth, $h_c$, which is the depth of the indentation that remains after the unloading is complete . To determine the critical depth of indentation from the load-displacement curve, a slope is drawn at the beginning of the unloading phase. This slope intersects the x-axis at the position corresponding to the critical depth, denoted as $h_c$. The critical depth is then calculated using the equation P = Sh + C, where P represents the load (set to 0), S is the slope, $h_c$ signifies the critical depth, and C is a constant term. By accurately determining the critical depth, the subsequent calculation of the critical contact area becomes feasible.

Where, R is the radius of the indenter.

Now the hardness (H) of the materials can be determined using the following relation:

$$H = \frac{P_{max}}{A_c}$$

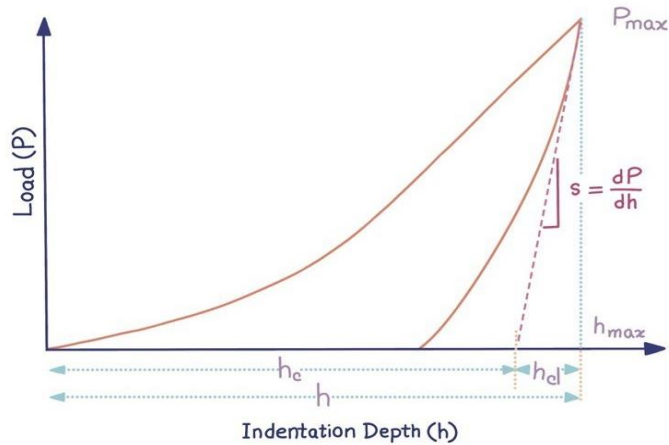

**Figure 2.6**: Schematic load-displacement curve (p-h curve)

## Contact atoms method

In a later study, Bolshakov and Pharr [47] proposed that in situations involving nanoindentation of sink-in materials or instances where material pile-up is prevalent (particularly when the ratio of residual indentation depth to maximum indentation depth (hc/hmax) surpasses the threshold of 0.7), there is minimal work hardening observed. These circumstances lead to an underestimation of the actual contact area by as much as 60% when employing the conventional approach of Oliver and Pharr. In our work the hc/hmax ratio was found to range from .7 to .9 (which is discussed in detail later). The findings indicate that utilizing the conventional approach of Oliver and Pharr may lead to overestimated hardness values when examining the plots. To mitigate this problem, another parameter referred to as "contact atoms" proposed by Shangda and Fujiu [46] was used. Shangda and Fujiu's method is extremely straight-forward and can be easily applied to MD simulation-based exploration of nanoindentation. This method, which aligns with our simulation approach of considering the spherical indenter as a rigid body, approximates the projection of all contact atoms to a circular shape. By employing a simple expression, the immediate contact area can be easily determined.

$$A(c) = \pi(R+r_o)^2$$

Where, R is the radius of the indenter and $r_o$ is the cutoff radius between the indenter and the substrate material (0.2 nm here). Each peripheral atom on the indenter possesses an interaction range extending up to " $r_o$," where it experiences repulsive forces from the atoms of the indented specimen. Consequently, by combining this parameter with the indenter radius, the overall radius necessary for calculating the total contact area is obtained. This method straightforwardly elucidates that modifying the tip radius leads to a proportional adjustment in the contact area between the indenter and the substrate.

## 3. Method Validation:

Our modeled $Ni_{60}Cr_{21}Fe_{19}$ has a density function which is closely related to actual density function of Inconel-718 at different temperatures which was studied in a previous work [48]. To validate our EAM potential file and nanoindentation simulation three more validations were done.

At First, we modeled a $Ni_{60}Cr_{21}Fe_{19}$ pristine cube structure with 605000 atoms. Raised the temperature from 300K to 2700K. Change in total energy with temperature was plotted in figure 3.1 and the graph showed a steep change of slope indicating bond breaking initiation. The second change of slope dictates the complete melting of the cube. The melting process started at 1475 K and complete melting happened at 1750 K.

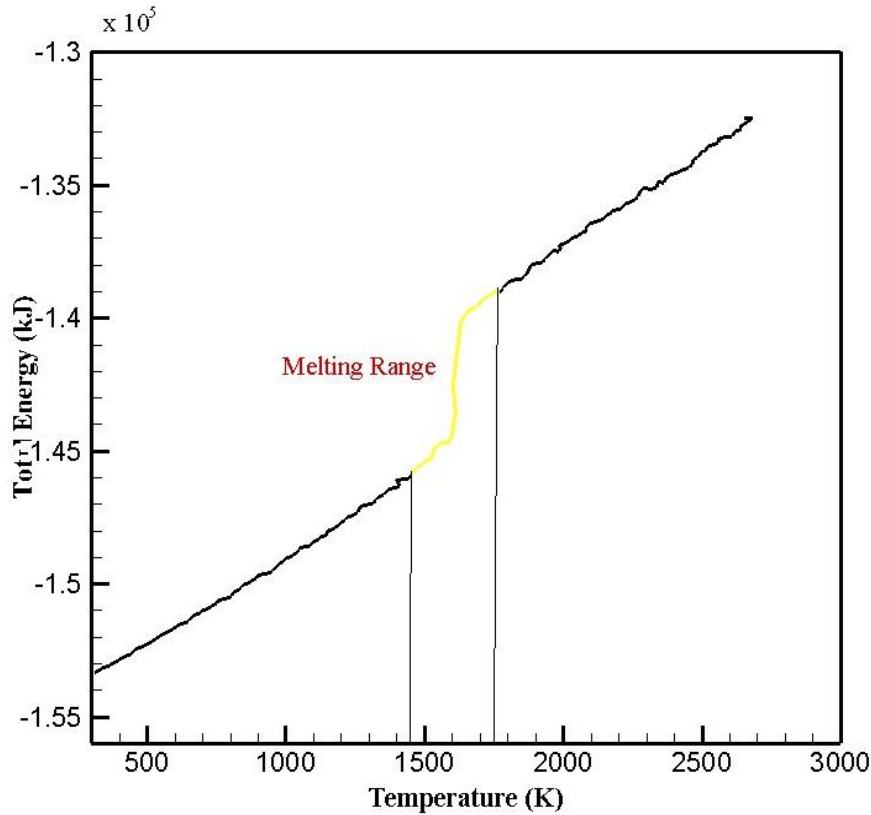

**Figure 3.1**: Change in total energy of Cube as Temperature increased from 300 K to 2700 K .

Experimentally available results indicate the melting of inconel at 1700 K which is in well agreement with our result. To validate our nanoindentation simulation we took an Al [001] oriented crystal structure and nanoindented it. The dimension of the structure(22.2 nm × 22.2 nm × 16 nm), potential file, indenter velocity(50m/s), indentation depth(1.5 nm) and indenter radius(5nm) were kept same as the study done by satyajit et al [49]. Our (P-h) curve showed well coherence with his work which is seen in figure 3.2

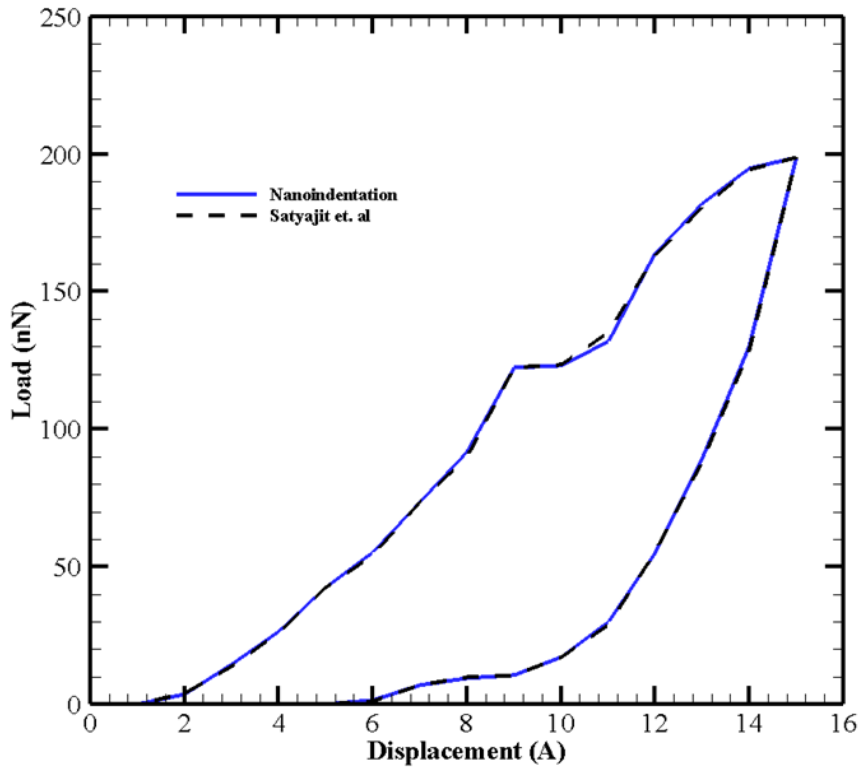

**Figure 3.2**: Lodaing-Unloading (P-h) curve according to our and satyajit et. al's Study.

In this study using contact atoms method, Hardness ranged between 3GPa – 7Gpa. Experimental results by H.Wang and A.Dhiman on AM Inconel-718 [50] suggested Hardness ranging from 4Gpa -5GPa which agrees well with this study.

# 4. Result and Discussion:

# Microstructure Evolution (Pristine):

| Cooling Rate (K/ps) | Microstructure Evolution | Solidified Layer |
|---|---|---|
| 5 | 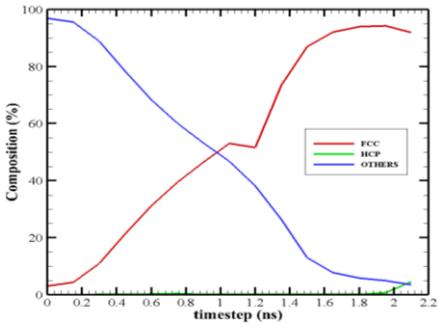 | 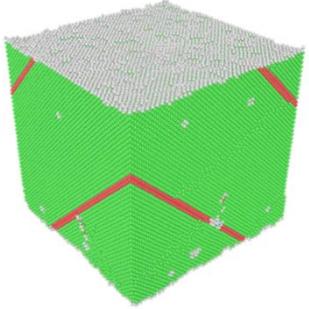 |
| 7.5 | 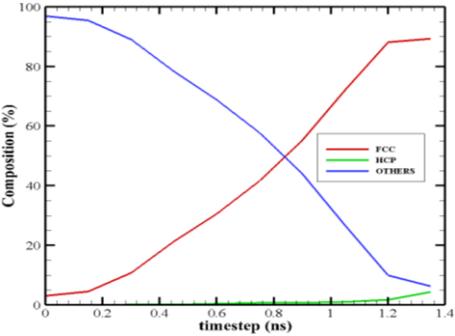 | 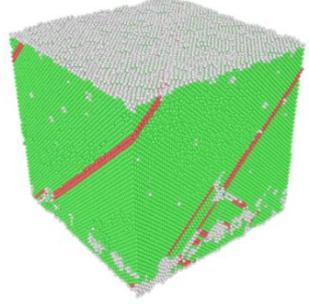 |
| 10 | 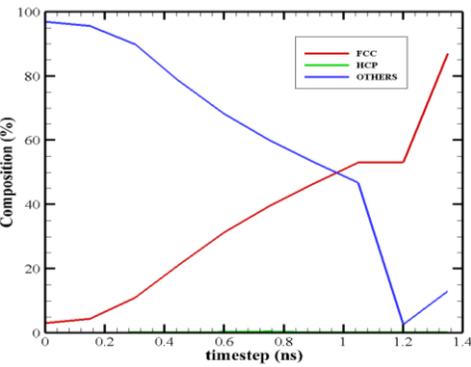 | 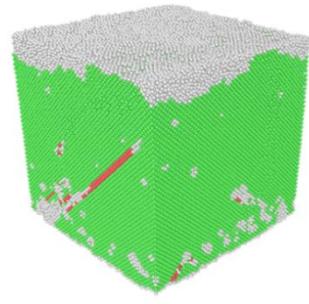 |

| 15 | 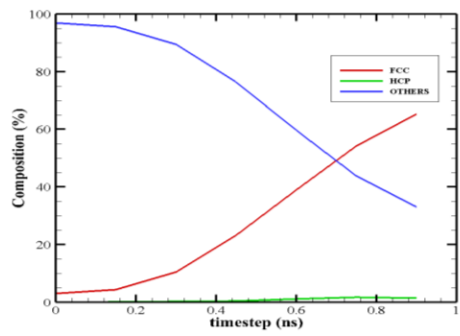 | 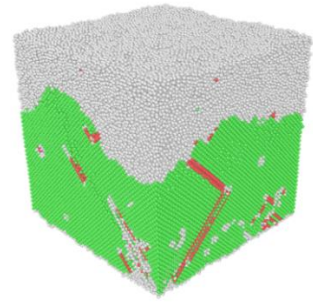 |
| 50 | 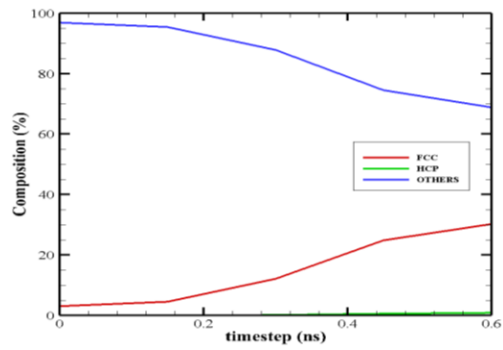 | 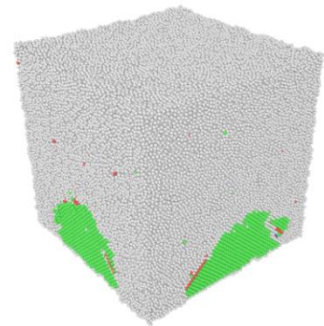 |
| 100 | 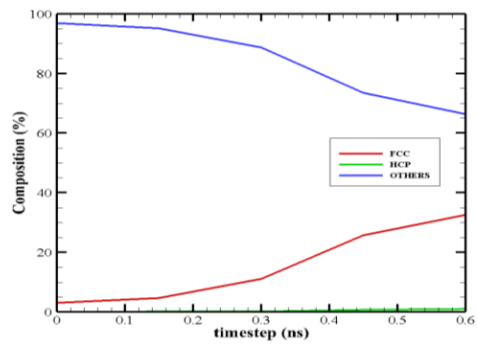 | 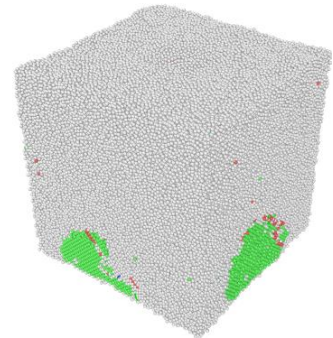 |

## Microstructure Evolution (Polycrystalline):

| Cooling Rate (K/ps) | Microstructure Evolution | Solidified Layer |
|---|---|---|
| 5 | 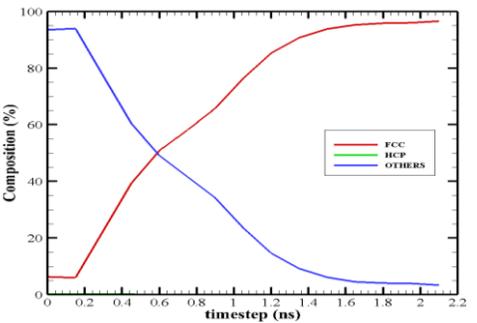 | 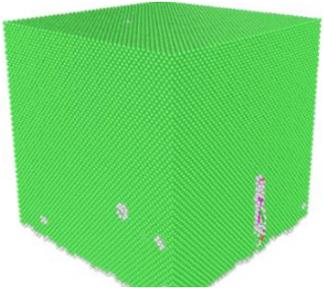 |
| 7.5 | 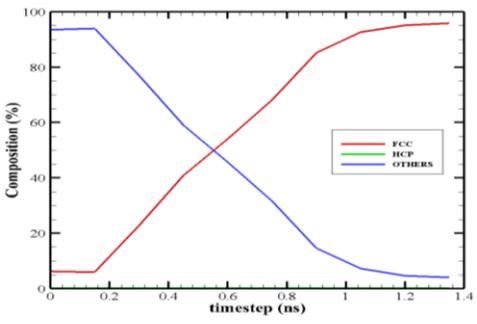 | 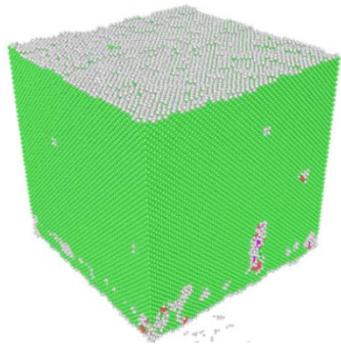 |

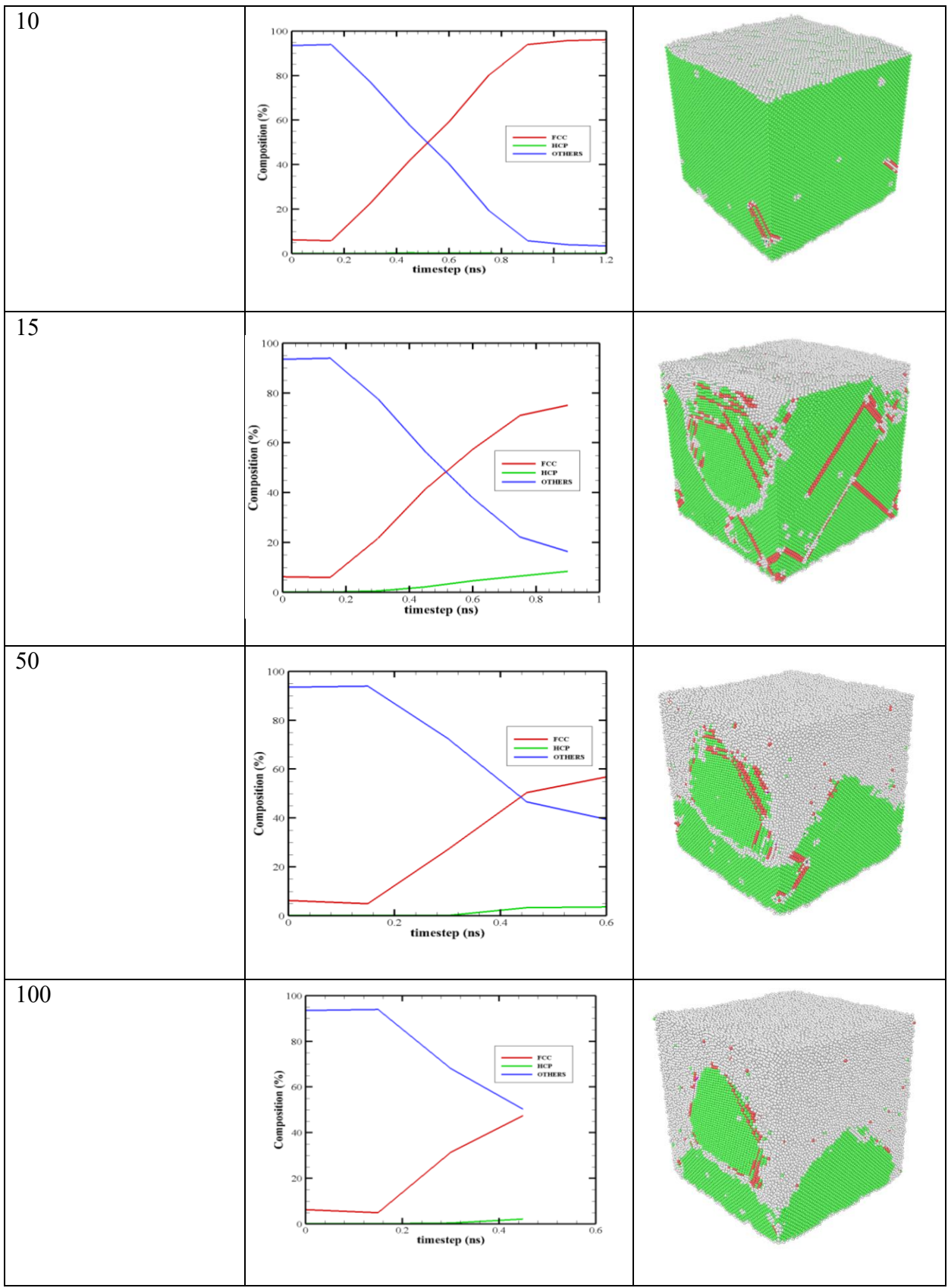

The effects of different cooling rates applied on the final microstructure of the material for pristine and polycrystalline is observed in this study. Here, lattice composition evolution has been observed for the AM simulation for various cooling rates. It is seen that the overall formation of FCC lattice of atoms is highest at lower cooling rates. However, in case of higher cooling rates, the formation of regular FCC stops completely after a certain point. That's why, there is high portion of amorphous phase present due to lack of nucleation for higher cooling rates. But in case of polycrystalline, grain boundaries can act as a barrier to dislocations and so, there is lack of residual dislocations, slip planes in the final structure results in higher FCC portions than pristine counterpart.

## 4.1 Effect of cooling rate on loading-unloading curve

Loading-unloading graph or P-h graph during nanoindentation at different cooling rates is plotted. As we start from, slower cooling rate, 5 K/ps, We can see the P-h curve shows a maximum load 182 nN in polycrystal structure but in pristine structure the value of maximum load is 157 nN. As the value of $h_c/h_{max}$ is .977 for polycrystal it shows more ductile behaviour in nature. In case of pristine structure the value of $h_c/h_{max}$ is .75 which indicates more brittle in nature. As we increase the cooling rate to 7.5 K/ps we can see an increase in maximum load, which is about 194 nN and 184 nN respectively for polycrystal and pristine structure.

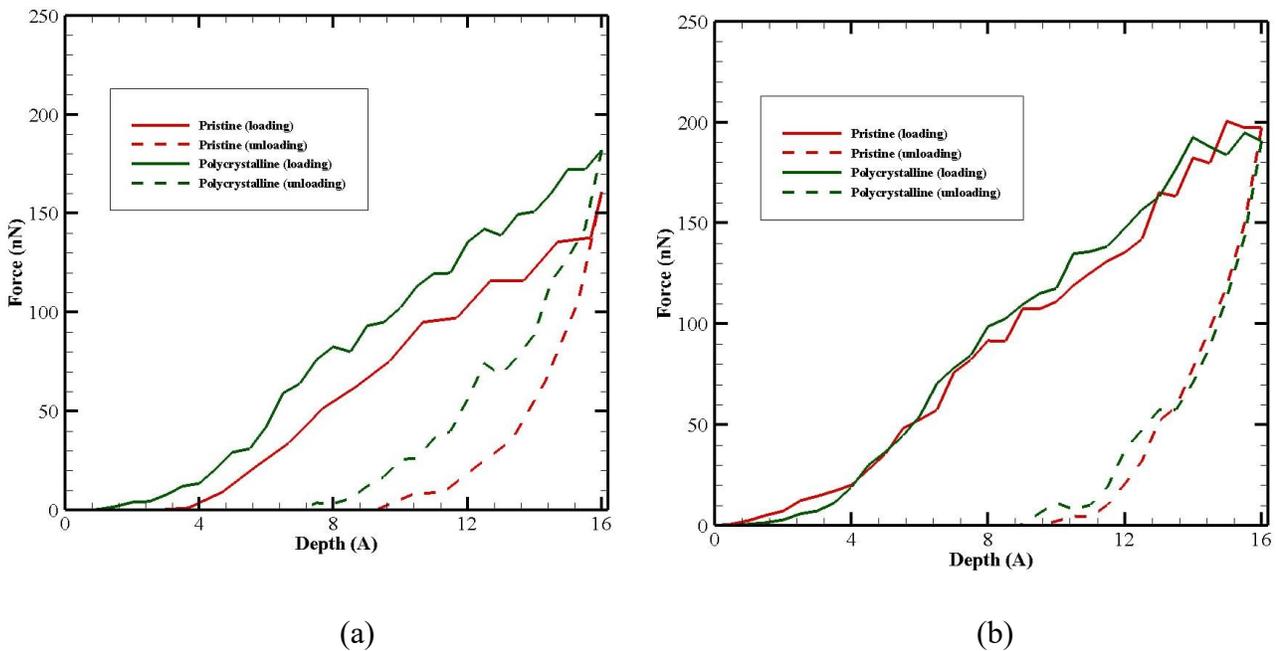

(a)           (b)

**Figure 4.1**: (a) 5 K/ps (P-h) curve   (b) 7.5 K/ps (P-h) curve

Both 10 K/ps and 15 K/ps show much higher maximum load for 1.6 nm indentation depth than other cooling rates which are about 205 and 189 nN for polycrystal and pristine structure. These two cooling rates show more refined microstructure, which can lead to an increase in the strength and hardness of the material. This can result in an increase in the Pmax value for the same depth of indentation, indicating that the material is able to withstand higher loads before undergoing plastic deformation.

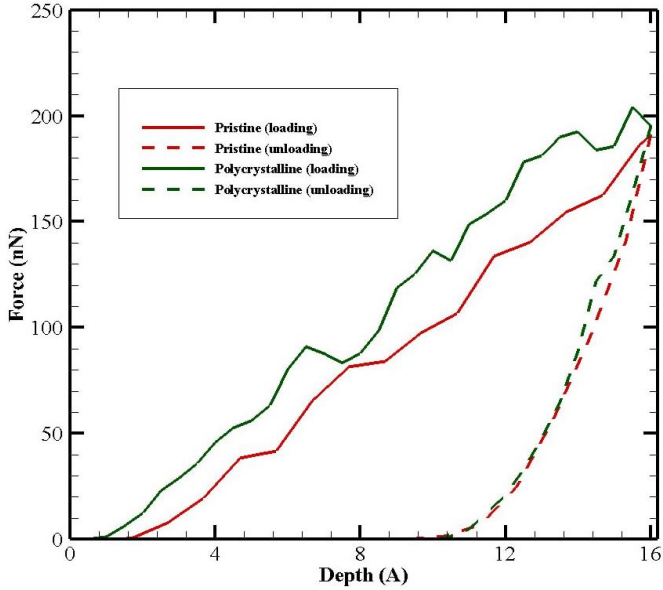

(c)

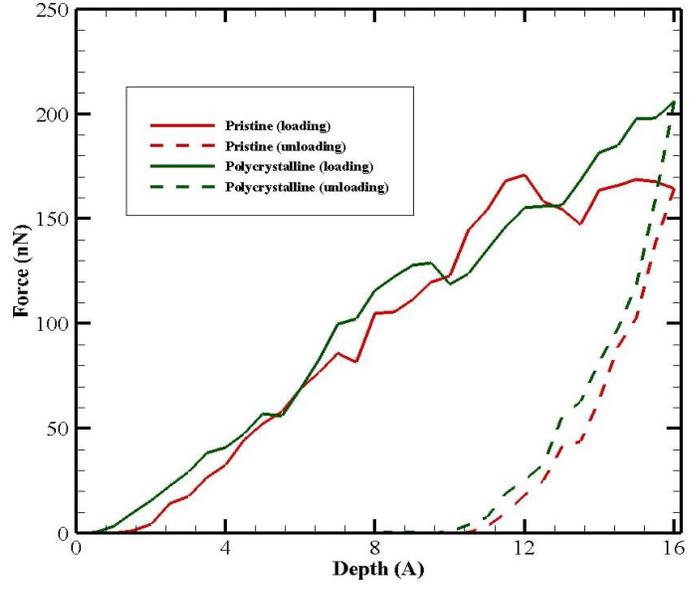

(d)

**Figure 4.2**: (c) 10 K/ps (P-h) curve   (d) 15 K/ps (P-h) curve

As the previous cooling rate results were indicating higher cooling rates were creating more refined structure with higher barrier to indentation. But for 50 K/ps and 100 K/ps, the maximum loads range from 90 Nn to 122 nN, significantly lower than previous cooling cases

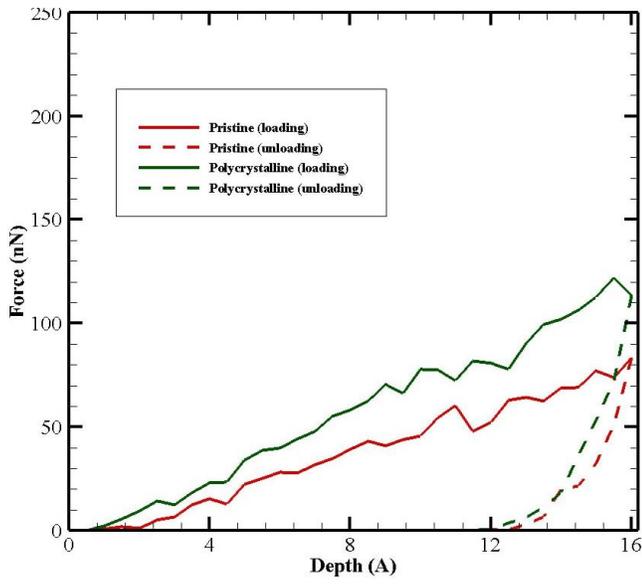

(e)

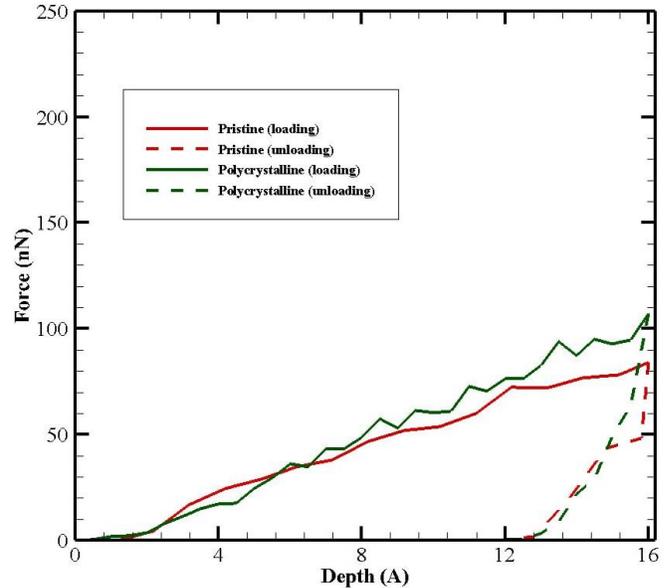

(f)

**Figure 4.3**: (e) 50 K/ps (P-h) curve  (f)100 K/ps (P-h) curve

We can conclude that rapid quenching can result in a poor microstructure and slower nucleation, which can affect the overall mechanical properties of the material being simulated.

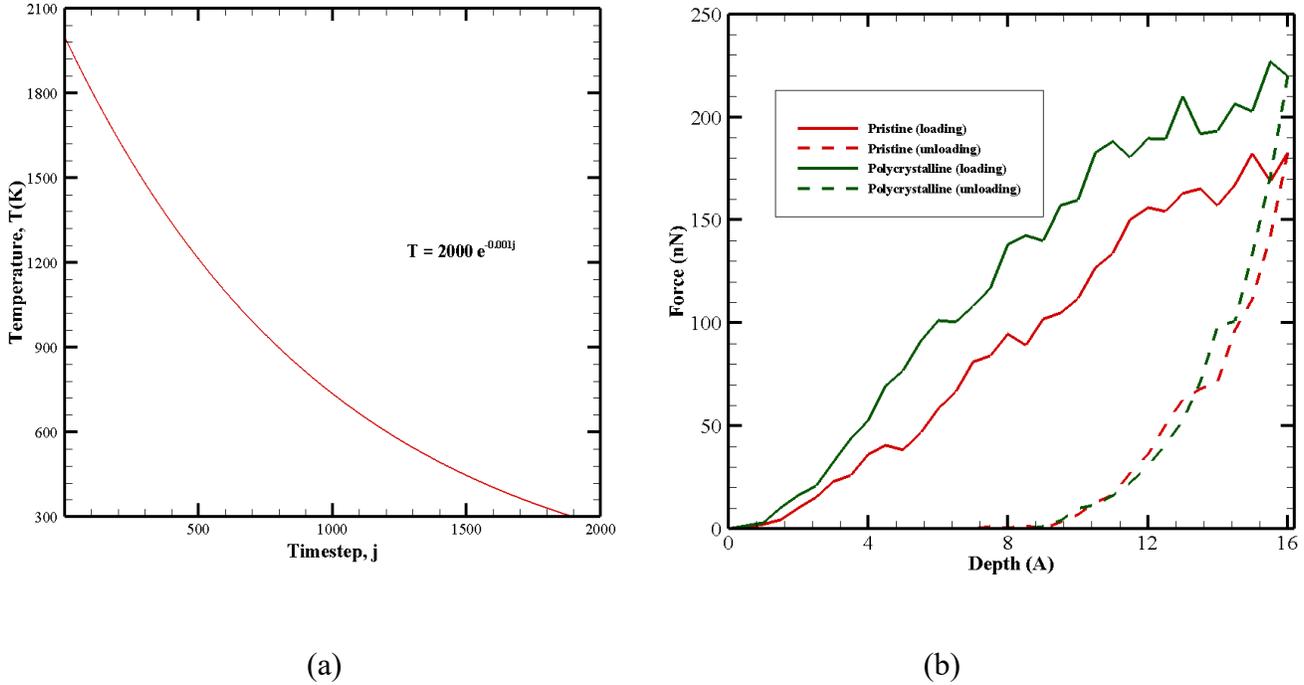

(a)           (b)

**Figure 4.4**: (a) exponential cooling profile (b) load displacement curve at expoinential cooling rate

All the previous cooling rates were constant. We introduced an exponentially cooled structure which showed higher loading force than constant cooling rates.

## 4.2    Effect of cooling rate on dislocation loops:

The nanoindentation simulation revealed the presence of dislocation loops, which are defects in the crystal lattice caused by the movement of dislocations. Dislocations are linear defects in the crystal structure that arise due to the mismatch between neighboring lattice planes. These dislocations can move through the crystal lattice under stress, causing plastic deformation.Figure 4.5 illustrates the formation and progression of dislocation loops under various cooling rates when loaded in the [001] direction. The findings reveal that as the material cools, different rates lead to distinct patterns of dislocation loop development. In Figure 4.6, the interaction between the indenter and the material during penetration is depicted. Specifically, when the indenter is pressed into the material, dislocations are generated along the primary slip plane, forming a pyramid-shaped structure. This structure can be described as a dislocation lock or tetrahedron. As the indenter continues to penetrate, at higher forces, the dislocation lock becomes disrupted, resulting in the emission of partial dislocations.

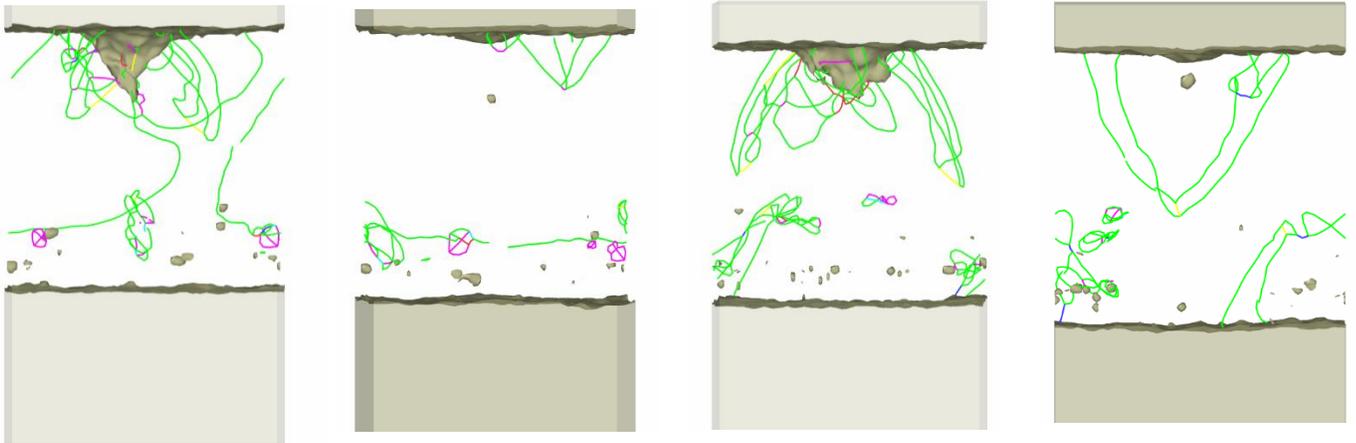

(a)5K/ps loading    (b) 5K/ps unloading    (c) 7.5K/ps loading    (d)7.5K/psunloading

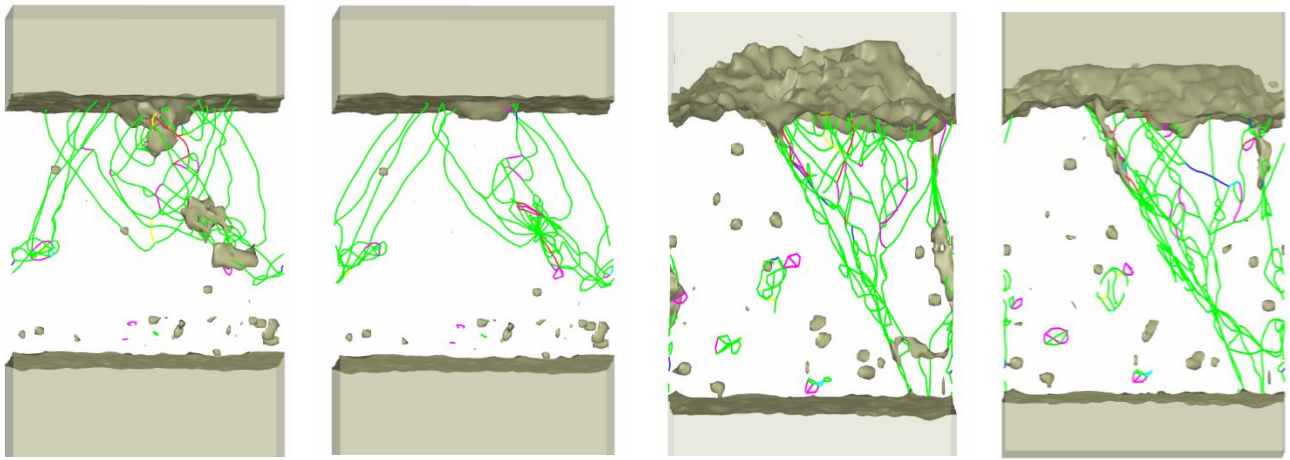

(e)10K/ps loading    (f) 10 K/ps unloading    (g) 15 K/ps loading    (h)15K/ps unloading

**Figure 4.5**: Dislocation loops and defect mesh of Pristine structure while loading and unloading at different cooling rates.

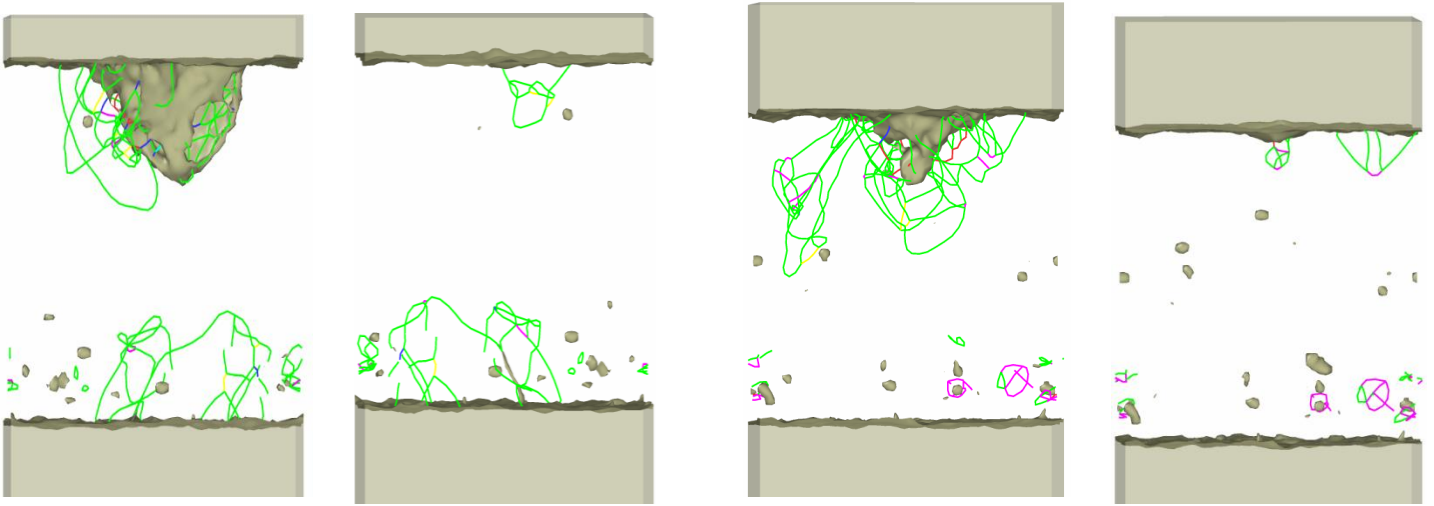

(i) 5 K/ps loading    (J) 5 K/ps unloading    (k) 7.5 K/ps loading    (l) 7.5 K/ps unloading

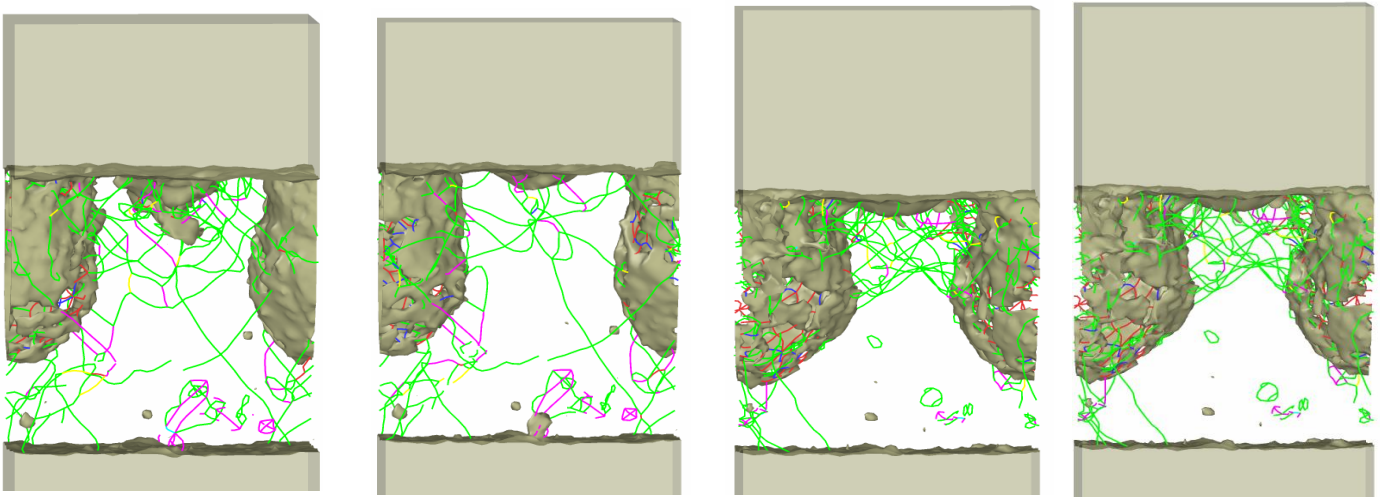

(e) 10K/ps loading    (f) 10 K/ps unloading    (g) 15 K/ps loading    (h) 15 K/ps unloading

**Figure 4.6**: Dislocation loops and defect mesh of Polycristalline structure while loading and unloading at different cooling rates.

Dislocation loops propagate much faster in higher cooling rates. Shockley partial type dislocation loops are mainly visible. We can also see the total dislocation lengths from Fig 4.7:

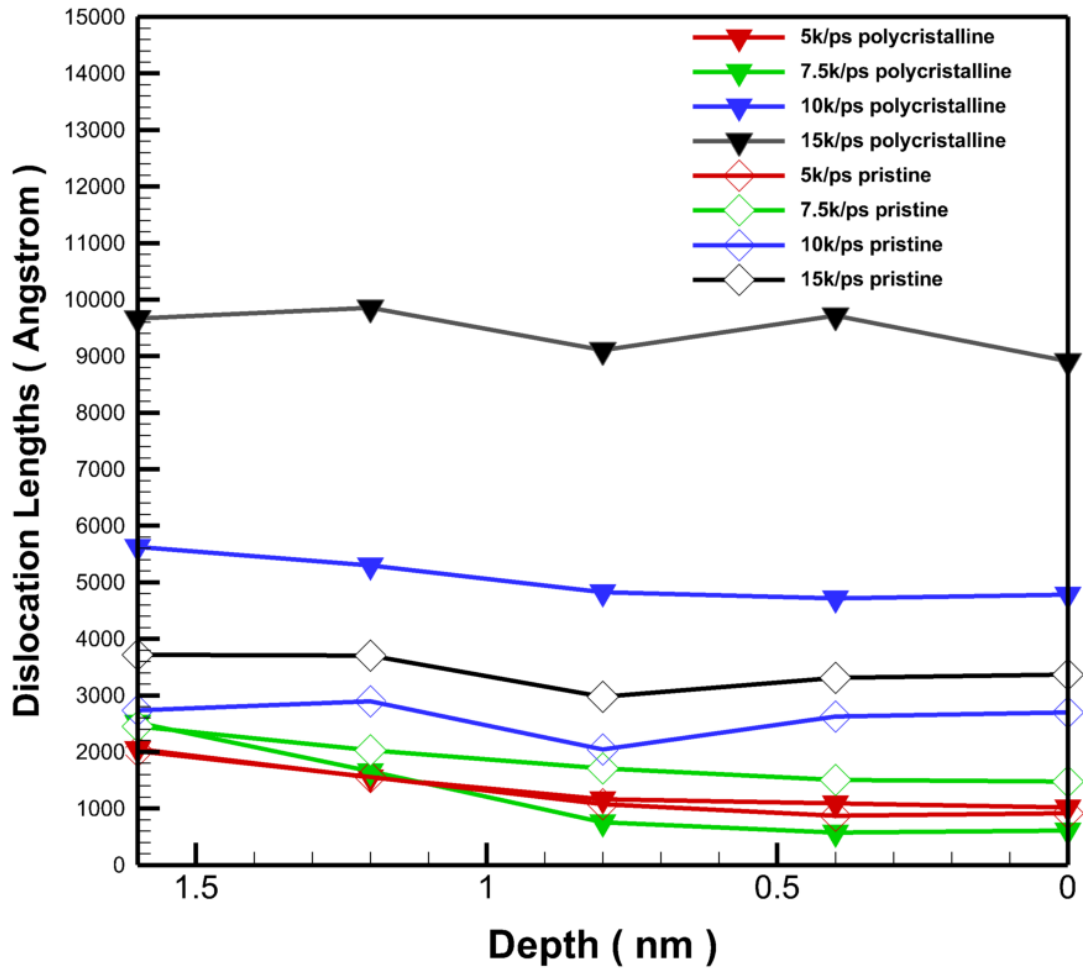

**Figure 4.7**: Total Dislocation lengths vs indentation depth at different cooling rate.

As the indenter reaches 1.6 nm maximum indentation depth the dislocations are showed for all the cooling rates. Then the indenter is unloaded and different dislocation lengths are found. Finally after total unloading is complete rest is the permanent dislocations inside the material. The study shows dislocations are increased with higher quenching rates.

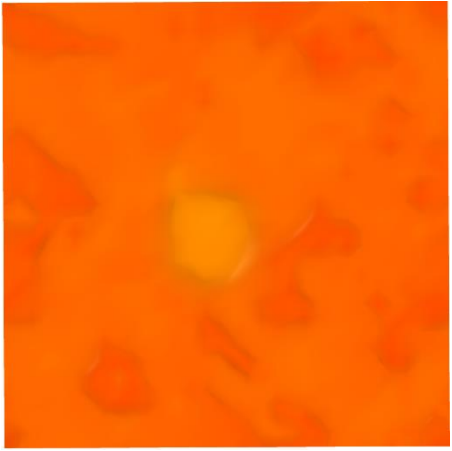 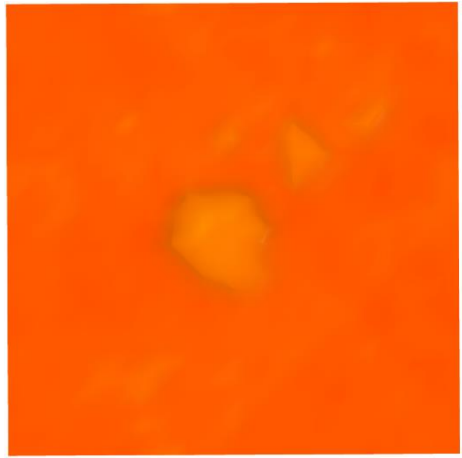 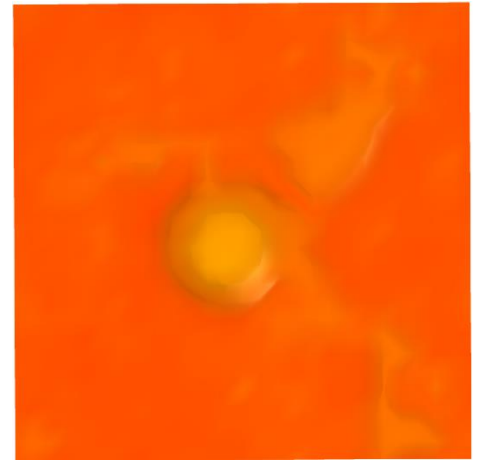

(a) 5K/ps  (b) 7.5K/ps  (c)10K/ps

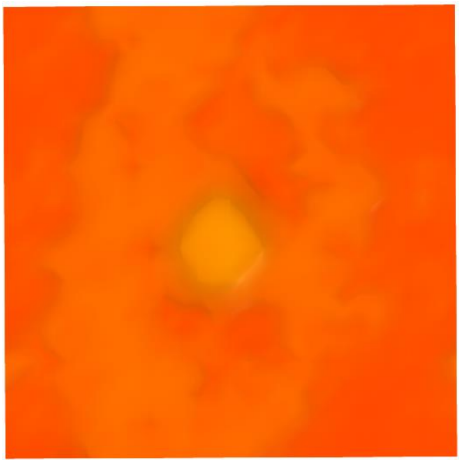 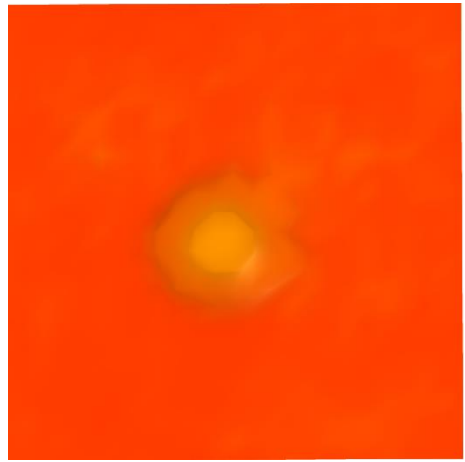 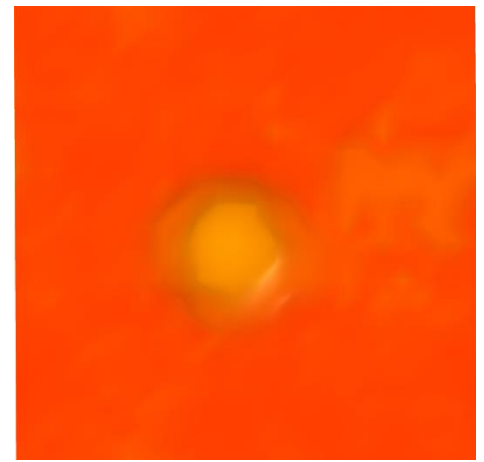

(d) 15K/ps  (e) 50K/ps  (f) 100K/ps

**Figure 4.8:** Surface imprint of pristine strucutre after unloading using Gaussian Density method.

The surface imprint that remains after the unloading process of indentation can be directly observed in experiments. This imprint plays a critical role in experiments, as it allows researchers to directly measure the critical depth of indentation from microscopic images of the imprint. Figure 4.8 shows surface imprints after indentation at different cooling rates. As can be seen, different cooling rates produce different types of surface imprints. With increased quenching, the surface imprint becomes more pronounced.

## 4.3 Hardness Calculation:

Before we calculate hardness, all the $h_c/h_{max}$ are given below:

Pristine Strucutre:

| Cooling Rate | $h_{max}$ | $h_c$ | $h_c/h_{max}$ |
|---|---|---|---|
| 5 | 1.6 | 1.213 | 0.758125 |
| 7.5 | 1.6 | 1.123 | 0.701875 |
| 10 | 1.6 | 1.174 | 0.73375 |
| 15 | 1.6 | 1.267 | 0.791875 |
| 50 | 1.6 | 1.285 | 0.803125 |
| 100 | 1.6 | 1.287 | 0.804375 |

Polycrystalline Structure:

| Cooling Rate | $h_{max}$ | $h_c$ | $h_c/h_{max}$ |
|---|---|---|---|
| 5 | 1.6 | 1.5643 | 0.9776875 |
| 7.5 | 1.6 | 1.2284 | 0.76775 |
| 10 | 1.6 | 1.2638 | 0.789875 |
| 15 | 1.6 | 1.3056 | 0.816 |
| 50 | 1.6 | 1.425 | 0.890625 |
| 100 | 1.6 | 1.4519 | 0.9074375 |

All the values of $h_c/h_{max}$ suggest that Oliver-Pharr method will bring great errors on hardness calculations. So, we calculated hardness according to Oliver-Pharr and contact atoms method, results are also plotted below.

The value of the hardness are shown in Fig 4.9 for different cooling rates. From the figure, it is evident that the hardness starts to increase when the cooling rate increases. Reaches maximum at 10 K/ps which is 6 GPa under Contact atom method and 10.857 GPa under Oliver-Pharr method. As we can see Oliver Pharr method is over estimating hardness because of pile-up problem.

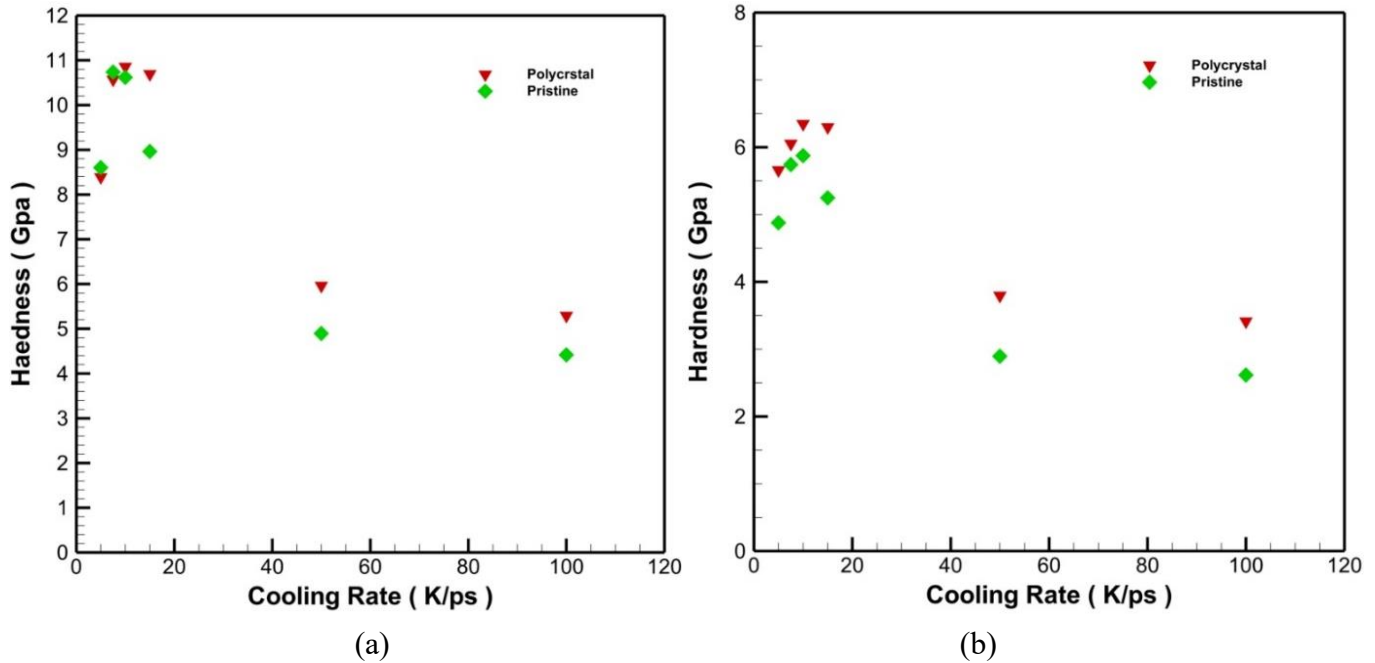

(a)                         (b)

**Figure 4.9**: (a) Effect of cooling rate on Hardness according Oliver Pharr method (b) Effect of Cooling rate on Hardness according to Contact Atom method

After 10 K/ps we can see a steep declination in hardness. At 10 K/ps, the atoms have sufficient time to rearrange and form a more ordered and stable microstructure. This results in a higher degree of strengthening, which is reflected in a higher hardness value during nanoindentation. However, as the cooling rate increases, such as 50 and 100 K per picosecond, the atoms have less time to rearrange and form a more ordered microstructure. As a result, the material is weaker and exhibits lower hardness during nanoindentation.

## Conclusion:

In this work, we used molecular dynamics (MD) to simulate the additive manufacturing (AM) process and nanoindentation of Inconel-718. Our results showed that:

- Very high quenching rates during AM can lead to low nucleation and poor microstructure.
- Increasing the cooling rate can increase hardness, but at very high cooling rates, hardness can decrease due to low nucleation. The optimum cooling rate in this study was 10 K/ps.
- The Oliver–Pharr method overestimates hardness in the presence of pileup, while the contact atom method is more accurate.
- Increasing the cooling rate increases the total dislocation length.
- Surface imprints are more prominent at higher cooling rates.
- Polycrystalline structures have higher hardness than single crystal structures.

The results and methods described in this work can be used to predict the nanomechanical behavior of AM metals. Additionally, this approach significantly reduces computation time and cost.